\documentclass[12pt]{iopart}
\usepackage{epsfig}

\newif\ifcomment
\newif\ifack
\newcommand{\dd}{\rm d}
\graphicspath{{./pics/}}

\begin{document}

\title[Recent Results from PHOBOS]{Recent Results from PHOBOS}

\author{C\ Loizides$^4$ for the PHOBOS collaboration\\
\vspace{0.2in}
%
%
%
B~Alver$^4$,
B~B~Back$^1$,
M~D~Baker$^2$,
M~Ballintijn$^4$,
D~S~Barton$^2$,
R~R~Betts$^6$,
A~A~Bickley$^7$,
R~Bindel$^7$,
W~Busza$^4$,
A~Carroll$^2$,
Z~Chai$^2$,
V~Chetluru$^6$,
M~P~Decowski$^4$,
E~Garc\'{\i}a$^6$,
N~George$^2$,
T~Gburek$^3$,
K~Gulbrandsen$^4$,
C~Halliwell$^6$,
J~Hamblen$^8$,
I~Harnarine$^6$,
M~Hauer$^2$,
C~Henderson$^4$,
D~J~Hofman$^6$,
R~S~Hollis$^6$,
R~Ho\l y\'{n}ski$^3$,
B~Holzman$^2$,
A~Iordanova$^6$,
E~Johnson$^8$,
J~L~Kane$^4$,
N~Khan$^8$,
P~Kulinich$^4$,
C~M~Kuo$^5$,
W~Li$^4$,
W~T~Lin$^5$,
C~Loizides$^4$,
S~Manly$^8$,
A~C~Mignerey$^7$,
R~Nouicer$^2$,
A~Olszewski$^3$,
R~Pak$^2$,
C~Reed$^4$,
E~Richardson$^7$,
C~Roland$^4$,
G~Roland$^4$,
J~Sagerer$^6$,
H~Seals$^2$,
I~Sedykh$^2$,
C~E~Smith$^6$,
M~A~Stankiewicz$^2$,
P~Steinberg$^2$,
G~S~F~Stephans$^4$,
A~Sukhanov$^2$,
A~Szostak$^2$,
M~B~Tonjes$^7$,
A~Trzupek$^3$,
C~Vale$^4$,
G~J~van~Nieuwenhuizen$^4$,
S~S~Vaurynovich$^4$,
R~Verdier$^4$,
G~I~Veres$^4$,
P~Walters$^8$,
E~Wenger$^4$,
D~Willhelm$^7$,
F~L~H~Wolfs$^8$,
B~Wosiek$^3$,
K~Wo\'{z}niak$^3$,
S~Wyngaardt$^2$,
B~Wys\l ouch$^4$}

%
%
%
%
\address{
$^1$~Argonne National Laboratory, Argonne, IL 60439-4843, USA\\
$^2$~Brookhaven National Laboratory, Upton, NY 11973-5000, USA\\
$^3$~Institute of Nuclear Physics PAN, Krak\'{o}w, Poland\\
$^4$~Massachusetts Institute of Technology, Cambridge, MA 02139-4307, USA\\
$^5$~National Central University, Chung-Li, Taiwan\\
$^6$~University of Illinois at Chicago, Chicago, IL 60607-7059, USA\\
$^7$~University of Maryland, College Park, MD 20742, USA\\
$^8$~University of Rochester, Rochester, NY 14627, USA}

\ead{loizides@mit.edu}

\begin{abstract}
In this manuscript we give a short summary of recent physics results from 
PHOBOS. Particular emphasis is put on elliptic flow, fluctuations in 
the initial geometry and the recent measurements of elliptic flow fluctuations. 
\end{abstract}


\submitto{\jpg}

\maketitle

\normalsize

\section{Summary of recent results}
In the first five runs~(2000--2005) of the Relativistic Heavy Ion 
Collider~(RHIC) at Brookhaven National Laboratory, the PHOBOS 
experiment~\cite{PHOBOS:nim} has collected data from 4 collision 
systems~(p+p, d+Au, Cu+Cu and Au+Au) in a wide range of collision energies~($19.6$, 
$22.4$, $62.4$, $127$, $200$ and $410$~GeV). To a large extent, analyses
of this dataset dealt with the global properties of charged particle 
production.
Recent results in this area extend the measurement of midrapidity charged 
particle multiplicity in Au+Au collisions to lower centrality, enabling a 
better overlap in $N_{\rm part}$ with the Cu+Cu system and include the 
analysis of the $\dd N_{\rm ch}/\dd\eta$ for all Cu+Cu energies, including the 
lowest energy data at $\sqrt{s_{_{\rm NN}}}=22.4$ GeV~\cite{Alver:2007we}. 
Furthermore, we have also obtained preliminary results on elliptic 
flow~\cite{Nouicer:qm06} in Cu+Cu collisions at $\sqrt{s_{_{\rm NN}}}=22.4$ 
GeV, completing our set of measurements on elliptic flow at RHIC for Cu+Cu
and Au+Au collisions at all available energies~\cite{PHOBOS:v2-130GeVAuAu,
Back:2004mh,Back:2004zg,PHOBOS:v2-epart}.
The final results for charged particle production
down to very low transverse momenta in Au+Au collisions at $\sqrt{s_{_{\rm NN}}}=62.4$ 
GeV~\cite{PHOBOS:spectra-auau}, as well as preliminary measurements of 
antiparticle to particle ratios~\cite{Veres:qm06} in Cu+Cu collisions have
been obtained.
A completely new analysis of charged two-particle angular correlations has been 
developed that fully utilizes the extensive coverage in pseudorapidity~($\Delta\eta 
\leq6.4$) and azimuthal angle~($\Delta\phi\leq2\pi$) provided by the 
PHOBOS multiplicity~(Octagon) detector and its corresponding ability to measure 
the full bulk of charged particle emission down to very low momentum. Studies 
related to this analysis initially have focused on the short-range correlations 
in pseudorapidity. Preliminary results for Cu+Cu~\cite{Li:qm06} and final results 
for p+p~\cite{Alver:2007wy} have recently been reported.

\section{Elliptic flow and eccentricity fluctuations}
\begin{figure}[t]
\begin{center}
\includegraphics[bb=0 20 555 348,width=0.8\textwidth]{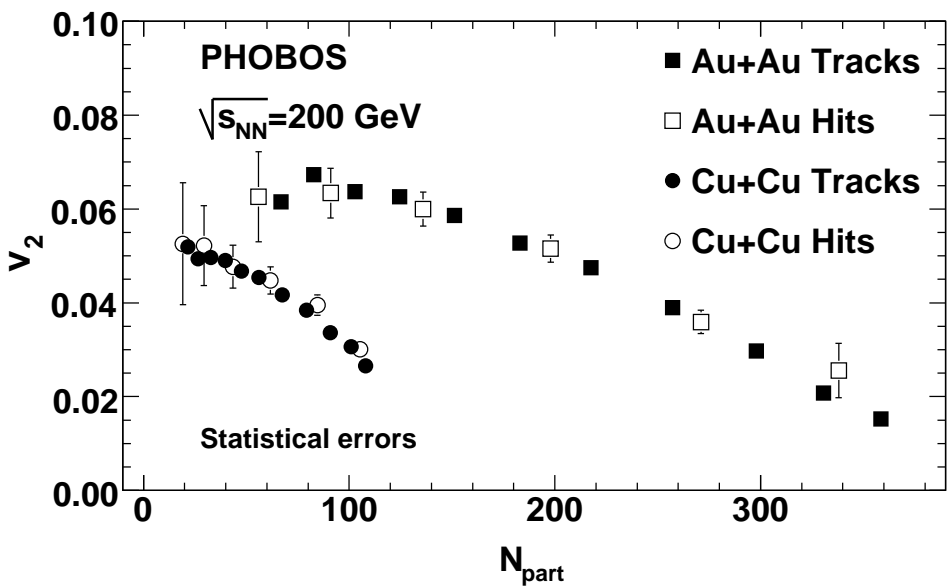}
\end{center}
\caption{\label{fig:v2vsnpart}
Magnitude of the average elliptic flow coefficient, $v_2$, 
at midrapidity as a function of centrality ($N_{\rm part}$) for Cu+Cu 
and Au+Au collisions at $\sqrt{s_{_{\rm NN}}}$ = 200 GeV.}
\end{figure}

Since the start of the RHIC program, studies of collective phenomena via the
measurement of the azimuthal distribution of produced particles have been one 
of the most important probes of the dynamics of nucleus--nucleus collisions. 
In particular, elliptic flow is sensitive to the early stages of the 
collision and its study provides unique insights into the properties of 
the hot, dense matter that is produced in these collisions.  
At the root of the interpretation of elliptic flow lies the connection to 
the initial overlap geometry of the colliding nuclei.
The azimuthal spatial asymmetry of the almond-shaped overlap region can only 
be reflected in the azimuthal distribution of detected particles if the produced
particles do significantly interact already shortly after the initial production.
At top RHIC energy, a large elliptic flow~($v_2$) signal near midrapidity has 
been observed~\cite{whitepaper}. Its magnitude is found to be in agreement with 
hydrodynamical model calculations of a relativistic hydrodynamic fluid 
which supports the current view that a strongly interacting state of matter 
is produced early in the collision process.
One of the primary motivations of colliding Cu+Cu and Au+Au nuclei at RHIC was 
to enable a detailed study of the effect of system size on all measurable physics 
observables. This is particularly interesting for elliptic flow. \Fref{fig:v2vsnpart} 
shows the magnitude of the elliptic flow, $v_2$, obtained from the PHOBOS hit-based
and track-based analyses,
as a function of centrality defined by the number of participants, 
$N_{\rm part}$~\cite{PHOBOS:v2-epart}. 
Two important features are immediately evident. First, the magnitude of flow in the 
smaller Cu+Cu system is large and qualitatively follows a similar trend with centrality 
as seen in the larger Au+Au system. Second, even for the most central collisions in 
Cu+Cu, the magnitude of $v_2$ is substantial, and exceeds that seen in central Au+Au
collisions.

\begin{figure}[t]
\includegraphics[width=0.4\textwidth]{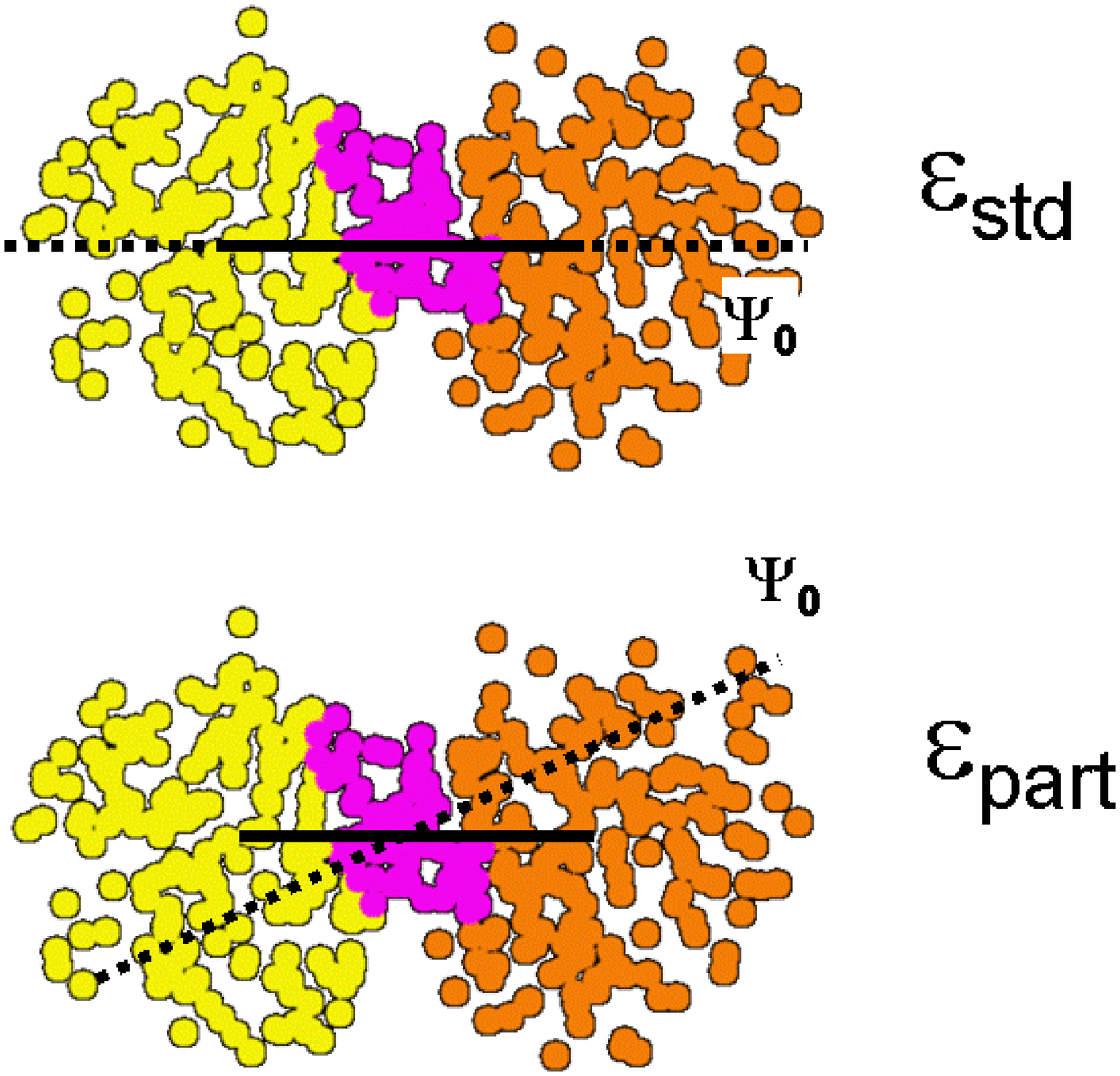}
\hspace{0.01\textwidth}
\includegraphics[width=0.59\textwidth]{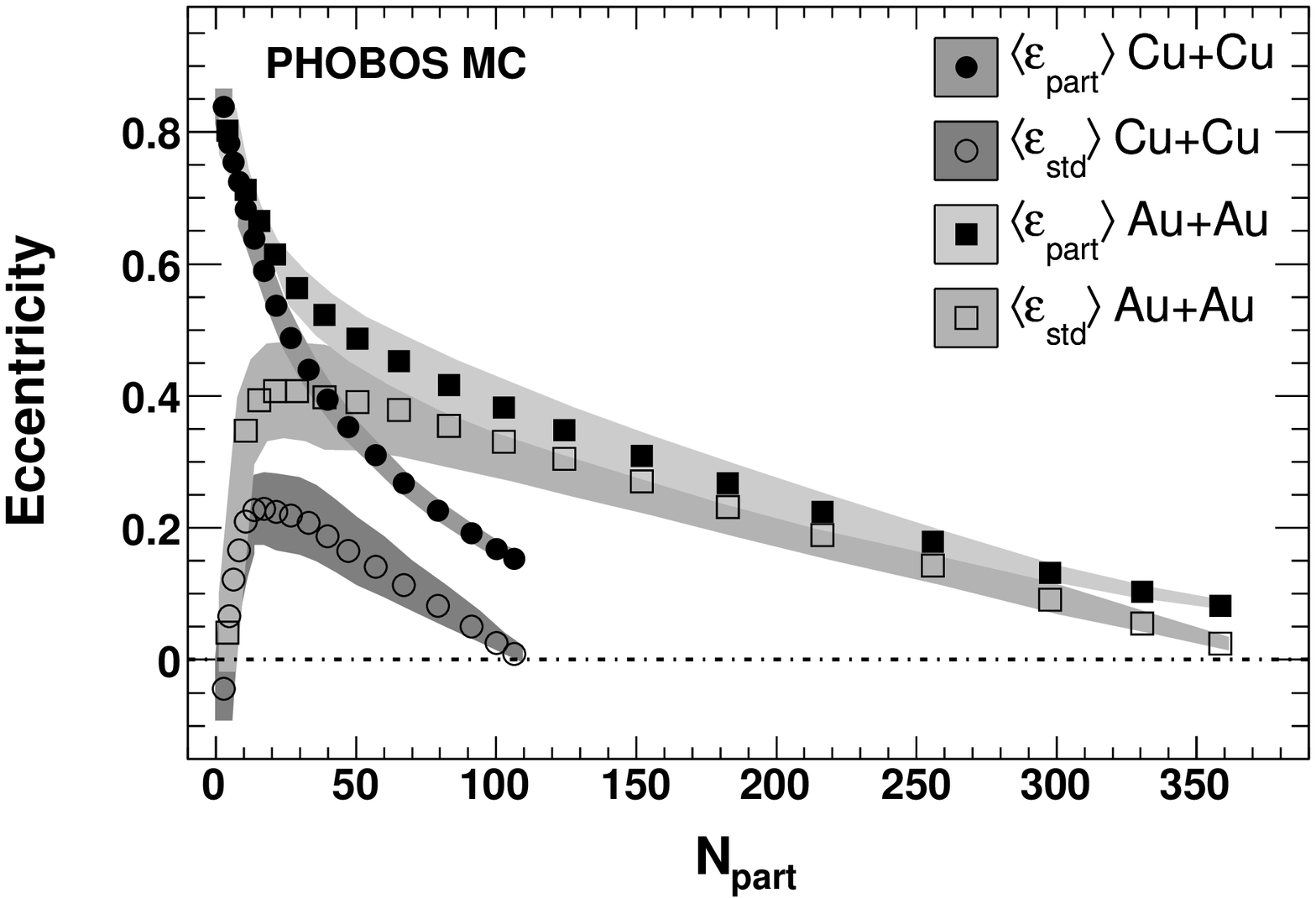}
\caption{\label{fig:eccdef}
Left: Visualization of the two approaches for calculating the 
eccentricity. The purple region (at center) in each collision illustrates
the interacting nucleons. The orange and yellow nucleons (away from
collision zone) are assumed not to directly influence the eccentricity.
The solid line represents the impact parameter direction, the dashed line
the direction of one of the axes in the calculation of the 
eccentricity~(rotated by $\Psi_0$ with respect to the nuclear reaction frame).
In the upper panel this direction is aligned with the impact parameter,
while in the lower panel it is aligned along the minor axis of the participant
region.  
Right: Comparison of $\langle\epsilon_{\rm std}\rangle$ and 
$\langle\epsilon_{\rm part}\rangle$ for Au+Au and Cu+Cu 
collisions at $\sqrt{s_{_{\rm NN}}}=200$ GeV. The grey bands show the systematic 
uncertainty from variation of the Glauber simulation parameters as described in 
Ref.~\cite{PHOBOS:v2-epart}.}
\end{figure}

\begin{figure}[t]
\includegraphics[bb=10 20 500 242,width=1.0\textwidth]{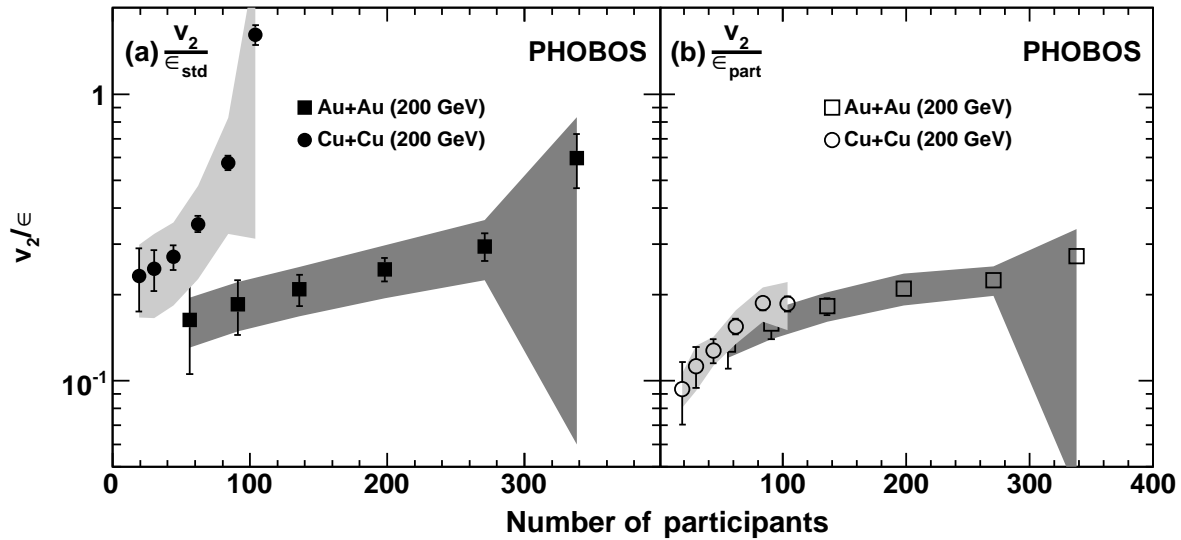}
\caption{\label{fig:v2scaled}
The average elliptic flow, $v_{2}$, scaled by the eccentricity from
a Glauber model calculation for the (a) standard and (b) participant
approaches.  Data are for Au+Au and Cu+Cu collisions at $\sqrt{s_{\rm NN}}=200$GeV.
Shaded bands (error bars) represent the systematic (statistical) uncertainty
from data.}
\end{figure}

In the most intuitive picture, the understanding of elliptic flow is
that the anisotropy of the azimuthal angle distribution of the final 
particles relative to the event plane is a consequence primarily of the 
initial eccentricity of the overlap region.
If this picture is correct, the elliptic flow results for both Cu+Cu
and Au+Au collisions for the same volume of the overlap region~($N_{\rm part}$) 
should be compatible if each is scaled by the proper eccentricity.
The eccentricity~($\epsilon$) of the overlap region can be estimated with 
the collection of participating nucleons. There are several definitions for 
calculating $\epsilon$, two of which are illustrated in \Fref{fig:eccdef}.
On the left, a schematic depiction of the ``standard'' (top, $\epsilon_{\rm std}$) 
and ``participant'' (bottom, $\epsilon_{\rm part}$) methods are shown. 
The former calculates the eccentricity of the overlap region assuming that 
the minor axis of the overlap region is aligned along the impact parameter.
The impact parameter and the beam direction define the {\it nuclear reaction 
plane}. However, fluctuations in the nucleon interaction points frequently create 
a situation where the minor axis of the overlap ellipse is not aligned with the 
impact parameter vector. The participant eccentricity definition~\cite{Manly:qm05}
accounts for this by quantifying the eccentricity with respect to the major axes 
of the overlap ellipse.
\begin{equation}
\epsilon_{\rm std}=\frac{\sigma^{2}_{y}-\sigma^{2}_{x}}
                    {\sigma^{2}_{y}+\sigma^{2}_{x}}
\label{eqn:eccstd}
\hspace{0.25\textwidth}
\epsilon_{\rm part}=\frac{\sqrt{(\sigma^{2}_{y}-\sigma^{2}_{x})^{2}+4\sigma^{2}_{xy}}}
                     {\sigma^{2}_{y}+\sigma^{2}_{x}}
\label{eqn:eccpart}
\end{equation}
Eqn~\ref{eqn:eccstd} is the mathematical representation of the eccentricity for 
both definitions, where
$\sigma_{xy}=\langle xy\rangle - \langle x\rangle\langle y\rangle$, 
$\sigma^2_{x}$ and $\sigma^2_{y}$ are the (co-)variances of the $x$ and $y$ 
participant nucleon position distributions expressed in the nuclear reaction 
frame. The difference in mean eccentricity between these two methods can be 
seen on the right-hand side of \Fref{fig:eccdef}. Deviations are clearly evident 
for peripheral and most central Au+Au collisions and for all centralities of 
Cu+Cu collisions, a result that illustrates the importance of finite-number 
fluctuations of the participant interaction points. 
This result is robust to the details of the Glauber Monte Carlo simulation,
as indicated by the bands which show the 90\% C.L. systematic errors.
\Fref{fig:v2scaled} compares the PHOBOS hit-based $v_2$ data scaled by 
$\epsilon_{\rm std}$ and $\epsilon_{\rm part}$. 
It is evident that the two very different systems are 
unified when scaled by the participant eccentricity. 
As recently presented, this unification 
when scaled by the participant eccentricity 
holds not only for the average value of $v_2$ at midrapidity, but also as a 
function of transverse momentum and pseudorapidity~\cite{Nouicer:qm06}.  

\begin{figure}[t]
\includegraphics[width=0.47\textwidth]{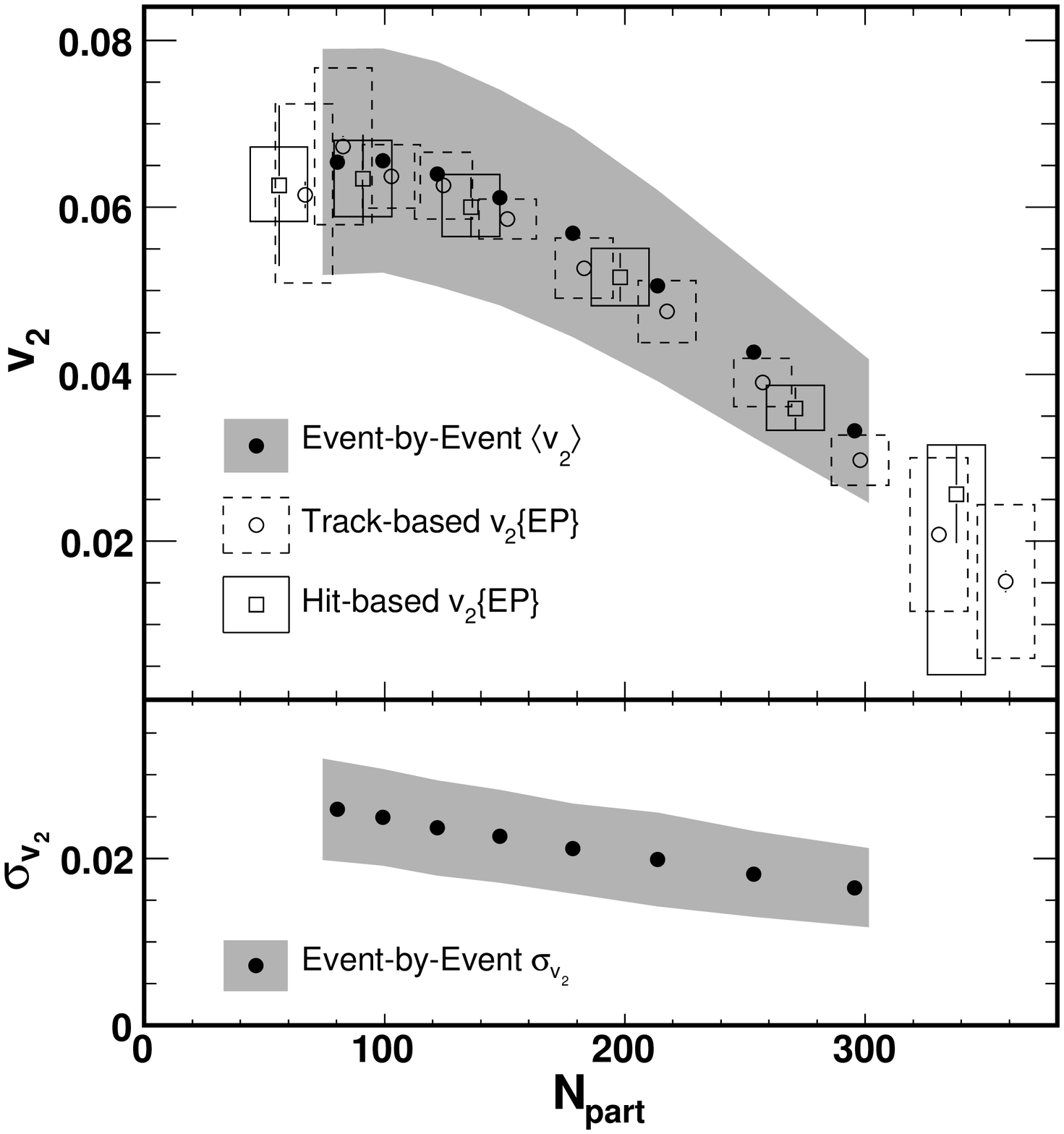}
\includegraphics[width=0.47\textwidth]{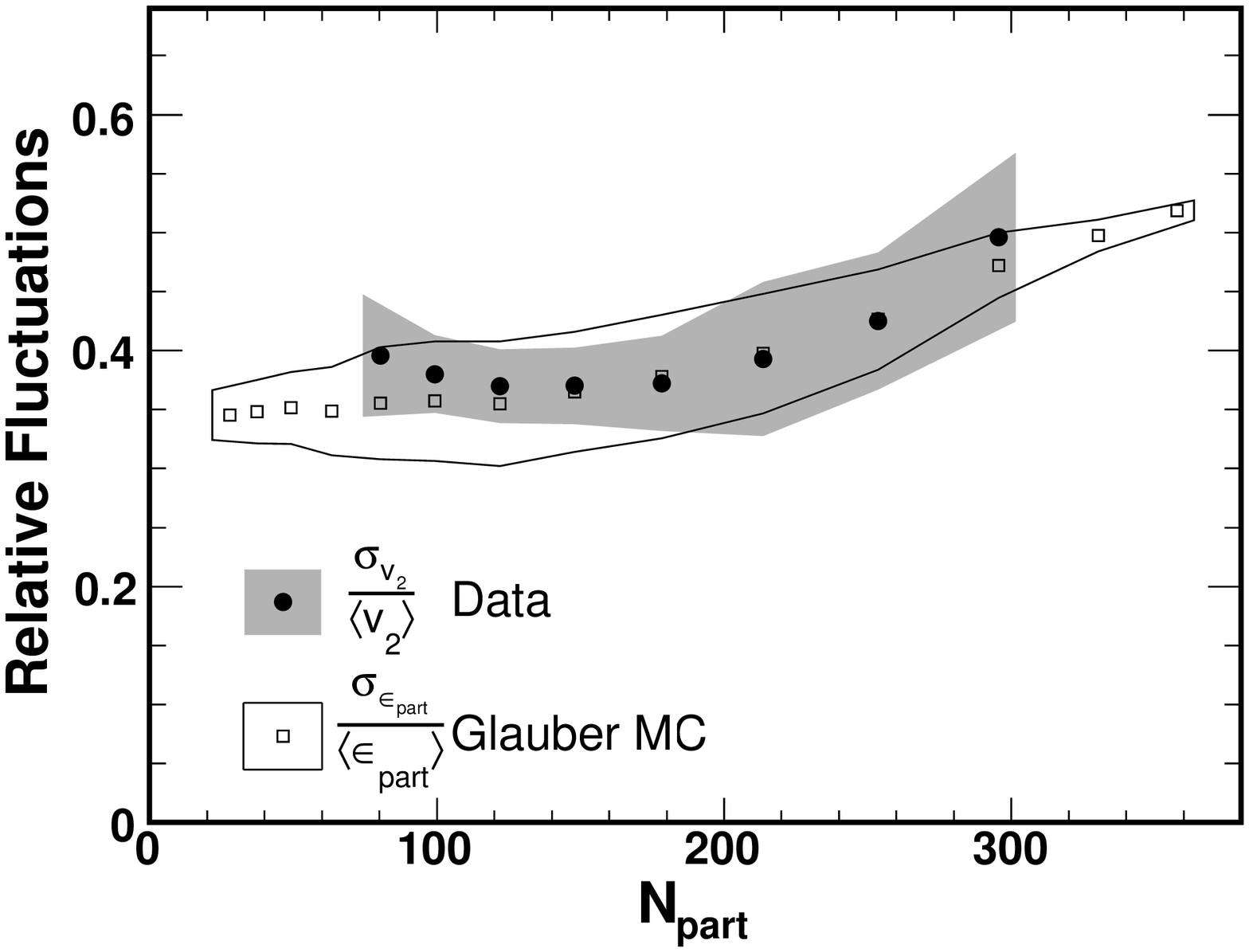}
\caption{\label{fig:resultmeanrms}
Left: Event-by-event measurement of $v_2$~(with corresponding $\sigma_{v_2}$ 
in the lower panel) at mid-rapidity compared with standard hit- and track-based 
PHOBOS results. Boxes and gray bands show 90\% C.L.~systematic errors and the 
error bars represent 1-$\sigma$ statistical errors. 
Right: Measured dynamical fluctuations in elliptic flow 
$\sigma(v_2)/\langle v_2 \rangle$, and participant eccentricity
fluctuations calculated in the PHOBOS participant eccentricity model.
The figures are taken from Ref.~\cite{Alver:2007qw}.}
\end{figure}

\section{Elliptic flow fluctuations}
The apparent relevance of the participant eccentricity model in unifying
the average elliptic flow results for Cu+Cu and Au+Au collisions leads
naturally to consideration of the dynamical fluctuations of both
the participant eccentricity itself as well as in the measured elliptic flow
signal from data. Simulations of the expected dynamical fluctuations 
in participant eccentricity as a function of $N_{\rm part}$
were performed using the PHOBOS Monte Carlo Glauber based participant 
eccentricity model, and they predict large dynamical fluctuations,
$\sigma(\epsilon_{\rm part})/\langle\epsilon_{\rm part}\rangle$ of the
order of 0.4 in Au+Au collisions at $\sqrt{s_{_{\rm NN}}} = 200$ GeV.
There are several different approaches one could develop to measure 
dynamical elliptic flow fluctuations. PHOBOS has recently
created a new method that is based on a direct measure of $v_2$ on 
an event-by-event basis using a maximum likelihood fit assuming that
the shape of $v_2$ in pseudo-rapidity is either triangular or trapezoidal
and that utilizes the unique large pseudorapidity coverage of the PHOBOS 
detector~\cite{Alver:ebye_v2_method,Alver:2007qw}. 
The strength of this approach lies in the fact that this analysis removes the 
effects of statistical fluctuations and multiplicity dependence by applying a 
detailed model of the detector response that enables both a measurement of the 
average $v_2$ on an event-by-event basis as well as a measure of the dynamical 
fluctuations in $v_2$.
The experimental results for both the average $\langle v_2 \rangle$ and the
measured dynamical fluctuations $\sigma_{v_2}$, which we also quantify using 
the ratio $\sigma(v_2)/\langle v_2 \rangle$, obtained in this new analysis are 
given in \Fref{fig:resultmeanrms} for Au+Au collisions at 
$\sqrt{s_{_{\rm NN}}}$ = 200 GeV~\cite{Alver:2007qw}.
The left-hand side of \Fref{fig:resultmeanrms} shows the results for the average
midrapidity elliptic flow obtained from the event-by-event analysis together with 
the results from both the standard hit-based and track-based analyses. 
The error bars represent statistical errors and the shaded bands the 90\% 
C.L.~systematic uncertainties. Confidence that all three measurements are determining 
the average elliptic flow is supported by the observation that they agree within 
the systematic errors. 
\enlargethispage{0.5cm}
The right-hand side of \Fref{fig:resultmeanrms} presents the PHOBOS results for $v_2$ 
dynamical fluctuations together with the result obtained for fluctuations in the 
participant eccentricity.  Systematics on the experimental measurement are decreased
by quantifying the result as a ratio of $\sigma(v_2)/\langle v_2 \rangle$.  We observe 
large dynamical fluctuations in elliptic flow with a magnitude in remarkable agreement
with calculations of participant eccentricity fluctuations.  
The observed agreement suggests that the fluctuations of elliptic flow primarily reflect 
fluctuations in the initial state geometry and are not affected strongly by the latter
stages of the collision. Note that the systematic errors imposed in our results include 
estimates from non-flow contributions to the observed magnitude of the flow fluctuations
that rely on a description of non-flow effects in HIJING. We are currently working on
an MC-independent way to estimate this contribution from data using two-particle
correlation measurements.

\section*{References}

\end{document}